\documentclass[]{spie}
\usepackage{natbib}
\usepackage{amstext}
\usepackage{amssymb}
\usepackage{amsmath}
\usepackage[amssymb]{SIunits}
\usepackage{latexsym}
\usepackage{placeins}
\usepackage{subfigure}
\usepackage{epsfig}
\usepackage{xspace}
\usepackage{moreverb}			
\usepackage{wrapfig}
\usepackage{fancyhdr}
\usepackage{ifthen}
\usepackage{lastpage}

\newcommand{\ArticleType}{paper\xspace}

\newcommand{\ie}{i.e.{}\xspace}			
\newcommand{\etc}{etc.\xspace}			
\newcommand{\cf}{cf.{}\xspace}			
\newcommand{\perse}{\emph{per se}\xspace}

\newcommand{\yyear}{\ensuremath{\mathrm y}}

\newcommand{\parsec}{\ensuremath{\mathrm {pc}} }
\newcommand{\eV}{\ensuremath{\mathrm {eV}} }
\newcommand{\AU}{\ensuremath{\mathrm {AU}} }

\newcommand{\sun}{\ensuremath{\odot}}

\newcommand\bra[1]{\ensuremath{\left\langle#1\right|}}
\newcommand\ket[1]{\ensuremath{\left|#1\right\rangle}}
\newcommand\iprod[2]{\ensuremath{\langle #1 | #2 \rangle}}

\newcommand\definedas{\ensuremath{\stackrel{\triangle}{=}} }

\DeclareMathOperator{\sinc}{sinc}

\DeclareMathOperator{\ambig}{ambig}

\newtheorem{corollary}{Corollary}[theorem]
\newtheorem{principle}{Principle}

\newcommand{\Section}[2]{\section{#2}\label{s:#1}}

\newcommand{\LI}{\begin{itemize}}
\newcommand{\IE}{\end{itemize}}
\newcommand{\LN}{\begin{enumerate}}
\newcommand{\NE}{\end{enumerate}}
\newcommand{\LD}{\begin{description}}
\newcommand{\DE}{\end{description}}
\newcommand{\LL}{\begin{list}}
\newcommand{\LE}{\end{list}}

\newif\iffigavailable
	\def\figavailable{\figavailabletrue}
	
	\figavailable

\begin{document}
\title{%
	Prediction of
		spectral shifts proportional to source distances
	by
		time-varying frequency or wavelength selection
}
\author{V.~Guruprasad
	\skiplinehalf
	Inspired Research, Brewster, New York, USA.
}
\maketitle
\begin{abstract}
Any frequency selective device with an ongoing drift will cause
observed spectra to be
	variously and simultaneously scaled
in proportion to
	their source distances.
The reason is that
	detectors after the drifting selection
will integrate
	instantaneous electric or magnetic field values
		from successive sinusoids,
and these sinusoids would differ
	in both frequency and phase.
Phase differences between frequencies
	are ordinarily irrelevant,
and recalibration procedures at most correct
	for frequency differences.
With drifting selection, however,
	each integrated field value comes from
		\emph{the sinusoid of
			the instantaneously selected frequency
		at
			its instantaneous received phase},
hence the waveform constructed by the integration
	will follow
		the drifting selection
	with a \emph{phase acceleration} given by
		the drift rate
	times
		the slope of
			the received phase spectrum.
A phase acceleration is literally
	a frequency shift,
and
	the phase spectrum slope of a received waveform is
		an asymptotic measure of the source distance,
as the path delay presents
	phase offsets proportional to
		frequency times the distance,
	and eventually exceeding
		all initial phase differences.
Tunable optics may soon be fast enough
	for realizing such shifts by Fourier switching,
and could lead to
	pocket X-ray devices;
	sources continuously variable from RF to gamma rays;
	capacity multiplication with
		jamming and noise immunity
	in
		both fibre and radio channels,
	passive ranging
		from ground to deep space;
	\etc

\end{abstract}
\pagestyle{fancy}
\Section{intro}{Introduction} 

Hitherto,
there has been no specific use or test of
	the continuity of the Fourier spectrum.
As stationary states,
even conduction band energy levels
	in metals and semiconductors
		are merely quasicontinuous.
The travelling wave energies are assumed to form a continuum
	in scattering and photoelectricity theories,
but only a quasicontinuous subset
	actually interacts with
		metallic or semiconductor states,
as represented by
	quantum interaction matrices of
		mixed cardinality.
The Doppler effect and the Lorentz transformation
	have been likewise formulated
		primarily in terms of
	their impact on individual pure tones
		present in the spectrum.
Since all observations are necessarily of finite precision,
specific use of spectral continuity
	requires involving a rate of
continuous drift of frequency or wavelength
	as a factor.
However,
though spectrometer drift is
	routinely corrected by recalibration
and has been especially well studied
	in astronomical instrumentation,
the observational impact of
	a continuously changing frequency or wavelength selection
		has not been specifically examined.
The principal result presented here is that
any frequency selective device with
	an ongoing drift of selection will cause
received spectra to be
	variously and simultaneously scaled
		\emph{in proportion to
			their source distances}
	to subsequent detectors.
This is remarkable as
the Hubble redshifts of astronomy are
	the only known instances of frequency shifts
		indicative of source distances.
As a general effect of
	tunable or ``Fourier switching'' optics,
it promises to add
	a whole new dimension in technology.

The reason is that
the subsequent detectors will integrate
	instantaneous electric or magnetic field values
from the successive sinusoids
	instantaneously presented by
		the drifting selector,
and
these sinusoids would differ
	in both frequency and phase.
Current recalibration procedures,
including
	those used for the space telescopes,
were designed to at most correct for
	frequency variation,
as integration across frequencies 
	was never considered.
Under a changing selection, however,
each successive integrated field value comes from
	\emph{the sinusoid of
		the instantaneously selected frequency
			at its instantaneous received phase}.
The waveform constructed by the integration
will therefore follow the changing selection
	with a \emph{phase acceleration}
defined by the rate of change times
	the slope of
		the received phase spectrum.
Phase acceleration defines
	frequency shifts,
and the phase spectrum slope is
	a direct asymptotic measure of
		the source distance,
since the total path delay presents
	phase offsets proportional to
		frequency times the distance,
and the offsets would eventually exceed
	all initial phase differences.
The shifts would be therefore
	proportional to
		individual source distances
in all observations with
	a changing frequency or wavelength selection!

The slowness of any uncorrected residual drifts
	and the enormous speed of light
together keep
	this frequency shift error undetectably small
		for ground sources.
However,
at the path delays and distances of cosmology,
	even drifts on geological time scales
should cause errors comparable to
	the Hubble redshifts.
Such drifts would occur from ordinary creep
	in metals and glasses,
which is currently uncorrected for
except under
	enormous stresses or high temperatures.
The shifts for
	local calibration referents,
used for both laboratory spectrometers
	and space telescopes,
would be undetectable
	owing to the small distance.
I show below that
	these shifts exactly fit
both the cosmological
	acceleration and time dilation,
and even satisfy Tolman's test
	\cite{Tolman1930,Hubble1935}
for 
	the reality of the expansion!

The very notion of exploiting phase differences
	across wavelengths or frequencies
		is generally new.
Holography, synthetic aperture imaging and interferometry
	all involve phase differences
		at individual wavelengths.
Pulse radar imaging exploits
	phase differences across the frequency comb
		arising from the pulse repetition,
but only indirectly
	via a Fourier inversion
	\cite{Prasad1986}.
The spectral phase profile must be characterized in
	ultrashort laser pulse technology,
but only for
	the subsequent pulse shaping use.
Optical devices are expected to soon reach tuning speeds
	sufficient to realize
		these shifts on earth,
and their validation would have
	immense theoretic and practical implications.

The distance proportionality makes them
	the general time domain analogue of spatial parallax.
They would enable 
	separation of signals from cochannel transmitters
		by transmitter distance
	and
		instant range determination
by physical means
	independently of the signal content,
in analogy to
	the known angular separation of incoming signals
	and
		the instant determination of transmitter bearing
	using directional antennae.
They would thus enable
	arbitrary band-limited communication
simultaneously between
	arbitrary pairs of locations,
by physics instead of
	modulation and multiplexing
		in time or frequency domains.
This would not only help realize
	Shannon's vision of
		simultaneous point-to-point communication
	\cite{Shannon1948},
but would
	multiply channel capacities 
by their concurrent reuse
	by noncolocated transmitters.
They could be used independently,
or in combination with
	all current technologies
	like TDMA, FDMA, CDMA or WDMA
		in air, space or optical fibres,
	for multiplying bandwidths.
The physical plane separation of signals by source distance
	would also endow receivers with
inherent immunity from
	in-channel noise and jamming.
They would additionally simplify
	passive radar and sonar
by obviating
	computation intensive path time correlations
		with multiple sources,
and enabling aperture synthesis
	without phase reference.

Another class of potential applications for
	the proportional shifts
is as
	frequency and wavelength transformers.
Transformers could be constructed
to turn
	radio, terahertz or infrared wavelengths
		into visible,
or
	visible
		into ultraviolet or x-ray
	wavelengths,
or the other way around,
	with accuracy and continuous tunability.
Wavelength transformation or scaling
	could be used
both for imaging and diagnosis,
	by scaling received wavelengths
		to values more suitable for
			observation and analysis,
and for synthesis or generation,
	by scaling the wavelengths of available sources
		to values desired in any particular application.
For example,
we might obtain
	coherent x-ray beams by scaling optical lasers,
or
	high power coherent infrared or optical beams
		by scaling up microwaves.



\begin{wrapfigure}[14]{l}{70mm}
	\centering
	\psfig{file=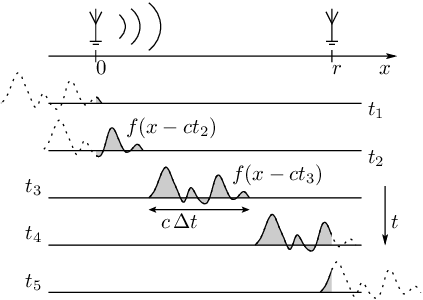, width=70mm}
	\caption{Travelling waveform}
	\label{f:travelwave}
\end{wrapfigure}
The rest of this \ArticleType is organized as follows.
The continuity of real travelling wave spectra
	is established in Section \ref{s:cont},
as we propose
	practical use of it
		for the first time.
Section \ref{s:beat} shows that
	the spectral continuity implies the presence of
		full path length information
			in received spectra
	in the form of
		the slope of phase spectrum.
Section \ref{s:shift} describes how
	this path length information
can be transformed
	into innovative spectral shifts
by continuous change of
	the instantaneous frequency or wavelength selections
		in a receiver or observing instrument.
Sections \ref{s:comm}-\ref{s:veri} discuss
	the applications of the shifts and
		a theoretical verification.
Broader implications including
	time dilation in the Doppler effect,
		and to cosmology
are briefly discussed
	in Section \ref{s:broad}.


\vspace{10pt}
\Section{cont}{Continuity of travelling Spectra} 

Transmission of information or energy by
	a travelling waveform
requires that
	the waveform begin
		at a definite instant at a source location
	and end
		at a later instant at a receiver location.
Fig.~\ref{f:travelwave}
	illustrates the idea.
A finite real-valued waveform $f(x - ct)$ is shown
starting at instant $t = t_1$ at a source
		located at the origin $x = 0$,
and propagating
	all the way to a receiver at $x = r$.
The waveform is shown at successive instants
	$t_1$, $t_2$, \dots, $t_5$
when it arrives
	at the receiver.
The ordinary notation for
	a travelling waveform as
		$f(x \pm ct)$
does not expressly limit its existence to
	a total interval of $r/c + \Delta t$,
where $r$ is
	the distance between the source and the receiver
and $\Delta t$ denotes
	the entire duration of the waveform
		at the source or at the receiver.
Without this express limitation,
the waveform appears to continue to exist 
	for $t < t_1$ and $t > t_5$.
The usual portrayal of signals as
	time-varying functions
also makes it hard to notice that
	a travelling waveform is equivalently
a spatial function $f(x)$ with
	a finite compact support $[0, c \Delta t]$.

In physical terms,
	a discontinous spectrum means
that either
	the spectrum contains intervals
		of zero amplitude,
or
	has abrupt changes in amplitude,
so that either 
	the amplitude slope or the phase slope
		is undefined or diverges to $\pm \infty$
	at one or more points.
The continuity is needed for
	the existence of the phase spectrum slope,
on which
	the innovative method of Section \ref{s:shift} depends,
and is assured by
	the following basic result from Fourier theory.

\begin{lemma}[Spectral continuity] \label{t:cont} 
	The Fourier spectrum of
		any waveform of finite duration or extent
	must be continuous and infinite.
\end{lemma}

\begin{proof}
	We can impose
		a finite compact support $[0, T]$
	upon
		an arbitrary waveform $g(x)$,
	by multiplying
		$g(x)$
	\begin{wrapfigure}[11]{l}{75mm}
		\centering
		\psfig{file=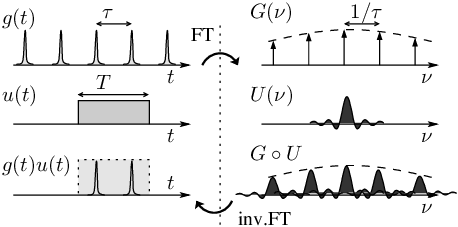, width=75mm}
		\caption{Spectral continuity}
		\label{f:fcont}
	\end{wrapfigure}
		with the interval function
	\begin{equation}
		u_{[0,T]}(x)
		\definedas 
			\begin{cases}
				1 &	\text{for~} x \in [0, T]
				\\
				0 &	\text{otherwise}
			\end{cases}
	\end{equation}
	and use the product
		$
		g(x) . u_{[0,T]}(x)
		$
	as a general candidate for
		a travelling waveform.
	Its spectrum would be
		the convolution product
		$
		G(\nu)
		\circ
		U_{[0,T]}(\nu)
		$,
	where
		$G(\nu)$
	is the Fourier transform of
		$g(x)$,
	and
		$
		U_{[0,T]}(\nu)
		$
	is the Fourier transform of
		$
		u_{[0,T]}(x)
		$,
	given by
		$
		U_{[0,T]}(\nu)
		\equiv
			2 T^{-1} e^{i T \nu / 2} 
			\,
			\sinc (2 \nu/T)
		$,
	where
		$\sinc(\xi) \equiv \sin(\xi)/\xi$
			is continuous and infinite.
	Even if $G(\nu)$ itself were discontinuous,
		say as a set of impulses,
	$U(\nu)$ would make
		the convolution product $G \circ U$
			continuous and infinite,
	as shown for
		a pulse train over time
		in Fig.~\ref{f:fcont}.
\end{proof} 

The requisite continuity of the phase spectrum
	is thus assured for travelling waveforms
on grounds that
	\emph{their finite duration makes
		their spectra continuous and infinite}.
For ranging applications,
it suffices to have
	a finite, well-defined phase spectrum slope
		at one or more frequencies.
For communication and wavelength transformation applications,
	the slope must exist
		over the entire spectrum,
barring
	an at most countable subset
		of frequencies,
whose contribution
	in the inverse Fourier transform
		would be vanishingly small.

In Fig.~\ref{f:fcont},
we have also accounted for
	the finiteness of real pulse widths,
which imposes
	an overall $\sinc$ profile (broken line)
		on the spectrum.
This outer profile limits the bandwidth
	in digital communication,
since a narrower bandwidth would widen pulses
	and reduce the bit rate.
The inner sinc profiles from
	the duration of transmission
have never needed
	consideration in the past
as they overlap
	the thermal linespreads
		of natural sources
and
	noise and clock jitter
		in communication.
The method of Section \ref{s:shift} extracts
	coherent information from these linespreads.


\Section{beat}{Source localization by differential path phase} 


Lemma \ref{t:sourceinfo} below explains how
	even strictly periodic waveforms
can transmit
	absolute source distance information.
This is ordinarily quite unintuitive,
and depends on being able to choose
	infinitesimally close pairs of wavelengths.
Lemma \ref{t:phaseslope} establishes more particularly that
	the slope of the phase spectrum gives
		a direct measure of the source distance,
and is the actual basis of
	the method of Section \ref{s:shift}.
The spectral continuity property
	established above
guarantees the existence of both
	infinitesimally close wavelength pairs
and
	phase spectrum slopes.

\begin{wrapfigure}[8]{l}{50mm}
	\centering
	\psfig{file=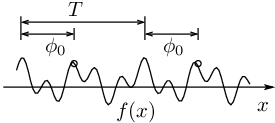, width=50mm}
	\caption{Periodic waveform}
	\label{f:periodic}
\end{wrapfigure}
Both lemmas are more generally applicable to periodic waveforms,
	which cannot of themselves
		represent travelling waveforms.
However,
decomposition by the Fourier transform leads to
	sinusoidal components
		which are periodic,
and other periodic functions are often used
	in signal processing.
The intuition in Lemma \ref{t:sourceinfo} is
that \emph{%
	the wavelength of the beat wave between
		infinitesimally close wavelength pairs
	is itself infinite,
which makes
	the beat wave aperiodic}.

Fig.~\ref{f:periodic} illustrates
how a known initial phase identifies 
	the possible locations of a wave source,
and thus partially determines
	its absolute location.
Consider a function $f(x)$
	which is periodic with the period $\lambda$,
\ie,
	$f(x + n \lambda) = f(x)$
	for integral values of $n$.
The generalized phase $\phi$ of $f$
	at an arbitrary point $x$ 
would be
	simply the scaled normalized value
	$
	\phi_f(x) 
	=
		2 \pi \, (x \bmod \lambda) / \lambda
	$,
relative to
	the origin,
which we may choose,
	without loss of generality,
as the location of
	our receiver.
If we determine the form of $f$,
	and its phase at the receiver location,
we would be able to compute $f$ 
	at any offset $x$ in the direction of propagation.
If $\phi_0$ denotes the known initial phase of
	the periodic component,
		in the overall decomposition,
then the set of the possible locations of the source
	is
	$
	\mathcal{L}_\lambda(\phi_0)
	=
	\{
		\dots ,
		(-1 + \phi_0/2\pi) \lambda,
		\phi_0 \, \lambda ,
		(1+\phi_0/2\pi) \lambda,
		(2+\phi_0/2\pi)\lambda,
		\dots 
	\}
	\equiv
	\{ (n + \phi_0/2\pi) \lambda
		,
		~
		n \in \mathbb{Z}
	\}
	$,
where
	$\mathbb{Z}$ is
		the set of integers.

\begin{wrapfigure}[10]{r}{50mm}
	\centering
	\psfig{file=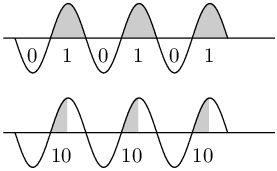, width=50mm}
	\caption{Localization information}
	\label{f:sineinfo}
\end{wrapfigure}
Another useful quantity is
	the density of $\mathcal{L}_\lambda$,
		which is simply $\lambda^{-1}$,
because the lower the density,
the less ambiguity we would have
	in locating the source
		within a given coordinate range.
For example,
if we knew by another means that
the source might be
	in the coordinate interval $[a, b)$,
the set of possible locations could be reduced,
	by a robust determination of
		the phase of $f$,
to
	$\mathcal{L}_\lambda \cap [a, b)$.
\emph{The ambiguity in localizing the source would be}
	$
	\lceil L / \lambda \rceil - 1
	$,
where $L \equiv |[a, b)|$ represents
	the length of the interval.
The source would be perfectly localized
	in the distance interval $[a, b)$
		if the ambiguity were zero.
Incidentally,
as accurate determination of phase
	is difficult,
and exact analogue comparison of
	a real valued quantity
		is in any case impossible,
we should more precisely represent
	the set of possible locations
in terms of
	$\phi_0 \pm \delta \phi/2$,
where $\delta \phi$ is 
	the uncertainty in the phase measurement.
However,
this uncertainty is actually inconsequential
	as the ambiguity comes from 
		repetitions of the wave period.
Consider that
	when $f$ is a sinusoid,
		$\phi_0 = 0$,
	and
		$\delta \phi = \pi$,
each positive (or negative)
	``lobe'' of the sinusoid
identifies
	an interval likely containing
		a point source
	(Fig.~\ref{f:sineinfo}, top).
If we improved the phase resolution to
	select a quarter cycle,
as shown at
	the bottom of Fig.~\ref{f:sineinfo},
we would have obtained $2$ bits of information to locate
	the source within each cycle,
but would not have identified
	its specific cycle.
\emph{The ambiguity in terms of localizing the source
		down to a specific cycle
	is thus independent of the precision 
		with which phase can be determined}.

Any periodic function would similarly have 
	an infinite ambiguity
in terms of localizing the source 
	to a specific cycle.
However,
an arbitrary linear combination of
	periodic or sinusoidal functions
need not be periodic.
If the periods $\lambda_1$ and $\lambda_2$ of
	a pair of periodic functions 
		are given to be rationally related,
\ie, there exist integers $a$ and $b$
	such that
		$\lambda_1 / \lambda_2 = a / b$,
then a combination of the two
	will have the period
		$a b \lambda$,
where $\lambda$ is defined by
	$\lambda_1 = a \lambda$
and
	$\lambda_2 = b \lambda$.
By the same reasoning,
a combination of functions of
	\emph{irrationally related} periods
		cannot be periodic.
Since rationally related periods constitute only
	a dense subset of the 2-D real number continuum $\mathbb{R}^2$,
the availability of
	absolute source distance information
from
	a truly continuous spectral decomposition
		cannot be ruled out!
However,
if extracting information from
	the \emph{nowhere dense} subset of
		irrational period pairs looks tricky,
the following trigonometric identities
\begin{equation} \label{e:trigbeat}
\begin{split}
	\sin (k_1 x) + \sin (k_2 x + \theta)
&
	=
	2
	\sin \left( \frac{[k_1 + k_2 ] x + \theta}{2} \right)
	\cos \left( \frac{[k_1 - k_2 ] x - \theta}{2} \right)
\qquad
	[ k_i \equiv 2 \pi / \lambda_i, i = 1,2 ]
\\
\text{and}
\quad
	\cos (k_1 x) + \cos (k_2 x + \theta)
&
	=
	2
	\cos \left( \frac{[k_1 + k_2 ] x + \theta}{2} \right)
	\cos \left( \frac{[k_1 - k_2 ] x - \theta}{2} \right)
\quad
	,
\end{split}
\end{equation}
show that
	combinations of sinusoids (of equal amplitudes)
		are \emph{always} periodic.
The solution is to use beat waves.


\begin{lemma}[Source localization] \label{t:sourceinfo} 
	Given a wavelength $\lambda$
	in the Fourier spectrum of
		a travelling waveform $f(x)$,
	and given
		a real number $L \gg \lambda$,
	a second wavelength $\lambda'$ exists
		in the spectrum
	such that
		the source can be located without cyclic ambiguity
			in the direction of travel 
		within any interval of length $|L|$
			containing the source,
	by measuring the phase of
		a combination of these Fourier components
			at an arbitrary distance from the source.
\end{lemma}

\begin{proof}
	From Fig.~\ref{f:sineinfo},
	the ambiguity of localizing to a cycle
		in the interval $L$
	with a sinusoid of period $\lambda$
		is
	\begin{equation}
		\ambig(L, \lambda)
		=
			\left\lceil
				L/\lambda
			\right\rceil
			-
			1
		\quad
			.
	\end{equation}
	By equations (\ref{e:trigbeat}),
	the beat wave
		between sinusoids of periods $\lambda$ and $\lambda'$
	would have the period
		$
		\lambda''
		\equiv
			|\lambda^{-1} - \lambda'^{-1}|^{-1}
		$,
	with the corresponding cyclic ambiguity
		$
		\ambig(L, \lambda'')
		$.
	We would therefore have
	\begin{equation} \label{e:longerbeat}
	\begin{split}
		\ambig(L, \lambda'') < \ambig(L, \lambda)
		\quad
	&
		\Leftrightarrow
		\quad
			\left\lceil
				L / \lambda
			\right\rceil
		<
			\left\lceil
				L / \lambda''
			\right\rceil
		\quad
		\Rightarrow
		\quad
			\lambda''
		>
			\lambda
	\\ &
		\Leftrightarrow
		\quad
			|\lambda^{-1} - \lambda'^{-1}|^{-1}
		\equiv
			\frac{
				\lambda' . \lambda
			}{
				| \lambda' - \lambda |
			}
		>
			\lambda
		\quad
		\Leftrightarrow
		\quad
			\lambda'
		>
			| \lambda' - \lambda |
	\quad
		,
	\end{split}
	\end{equation}
	\ie,
	we can reduce this ambiguity by simply choosing
		$\lambda'$ close enough to $\lambda$,
	and making sure
		the $\lambda'$ component has amplitude
			close to that of $\lambda$
	so that	
		equations (\ref{e:trigbeat}) can be applied.
	By Lemma \ref{t:cont},
	the spectrum of a travelling waveform must be
		both continuous and infinite.
	Hence,
	we can in principle choose $\lambda'$
		arbitrarily close to $\lambda$,
	and be also assured of
		equal amplitude.
	Doing so would make $\ambig(L, \lambda'')$
		arbitrarily small, including zero,
	since
	\begin{equation} \label{e:bestbeat}
		\ambig(L, \lambda'') < 1
		\quad
		\Leftrightarrow
		\quad
			1
		<
			\left\lceil
				\frac{L}{\lambda''}
			\right\rceil
		\quad
		\Rightarrow
		\quad
			\frac{
				\lambda' . \lambda
			}{
				| \lambda' - \lambda |
			}
		\ge
			L
	\quad
	\end{equation}
	\ie, close enough to $\lambda$,
		provided $L < \infty$.
	We would also need a way to determine
		the initial phase value $\phi_0$ of the beat wave.
	The continuity of the travelling waveform spectrum
		permits a reasonable assumption
	that
		the initial phases of adjacent sinusoids
			would be close as well.
	That is,
	we may assume the initial phase offset
		$\theta \rightarrow 0$
			in equations (\ref{e:trigbeat})
		as $\lambda' \rightarrow \lambda$.
	Then, regardless of what phases
		the sinusoids started with at the source,
	their initial phase difference,
		which gives the initial phase $\phi_0$
			of their beat wave,
		must be zero.
	A single measurement of the beat wave phase
		thus suffices to uniquely determine
			the source location within $L$.
\end{proof} 


Lemma \ref{t:sourceinfo} only establishes
	the existence of absolute source distance information
		in a travelling wave spectrum.
It is not suitable for practical use,
however,
as it calls for precise selection of 
	a pair of wavelengths $\lambda$ and $\lambda'$
that must be distinct but
	very close to each other.
Rearranging inequality (\ref{e:bestbeat}),
we get
\begin{equation}
	\lambda
	<
		\lambda'
		\le
			\lambda (1 - \lambda / L)^{-1}
		\approx
			\lambda (1 + \lambda / L)
\quad
	.
\end{equation}
The tightness of this bound is dictated by
	the useful choices for $\lambda$
		and expected range $L$.
For optical wavelengths
	$\lambda \sim 10^{-7}~\metre$,
even a laboratory range of
	$L \sim 1~\metre$
imposes a bound of
	$\lambda' \le \lambda (1 + 10^{-7})$.
This is certainly useless for
		terrestrial
	or
		astronomical distances,
as even for $L = 10~\kilo\metre$, 
we would need
	a wavelength resolution of $10^{-11}$.
This is complementary to
	the problem of measuring phase to some number of bits
		of precision in the first place,
as more than one bit of precision would be needed
	if we chose $\lambda$ comparable to $L$
to simplify
	the pair selection.

\begin{wrapfigure}[12]{r}{85mm}
	\centering
	\psfig{file=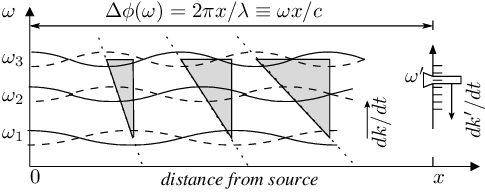, width=85mm}
	\caption{Spectral phase gradients}
	\label{f:phasegrad}
\end{wrapfigure}
An alternative to Lemma \ref{t:sourceinfo} comes from
	the theory of Green's functions,
which are solutions of
\begin{equation} \label{e:greendef}
	L G (x, s)
	=
		\delta(x - s)
\quad
	,
\end{equation}
where $L$ is a linear operator acting on $x$, 
	and $\delta(x-s)$ represents
		the Dirac delta.
When this is multiplied on both sides by $f(s)$
	and integrated over $s$,
the result is
\begin{equation*}
	\int L G (x,s) f(s) \, ds
	=
		\int \delta(x-s) f(s) \, ds
	=
		f(x)
\quad
	.
\end{equation*}
As
	$
	\int L G (x,s) f(s) \, ds
	\equiv
		L \int G (x,s) f(s) \, ds
	$,
this enables solution of
	$L u(x) = f(x)$
via
	$u(x) = \int G (x,s) f(s) \, ds$.
In diffraction theory,
boundary value problems are solved using
	$L = \nabla^2$,
		the Laplacian operator,
so that
	$\nabla^2 G(x,s) = \delta(x,s)$.
A general form of the boundary value problem
is given by
	Green's second identity
	$
	\int_V
		(
		\phi \nabla^2 \psi
		-
		\psi \nabla^2 \phi
		)
		\, dV
	=
	\int_S
		(
		\phi \nabla \psi
		-
		\psi \nabla \phi
		)
		\cdot
		d \vec{\sigma}
	$,
where $\vec{\sigma}$ is
	everywhere normal to the boundary surface $S$,
and is solved by setting
	$\psi = G$.
Trial solutions are applied of the form
	$\psi = s^{-1} e^{i k s}$,
where $s$ denotes
	distance from the source to the point of integration
		(see \cite[Chapter 8.3]{BornWolf}).
These trial functions are Green's functions
	corresponding to
		a point impulse source
and embody
	Fourier decomposition,
as $k \equiv 2 \pi / \lambda$ denotes
	the wave number of sinusoid.
The generality of this approach stems from the equivalence of
	an arbitrary source
		to a space-time distribution of point impulses,
so that
	the electromagnetic potential distribution and
		wave propagation
become
	integrals of Green's function solutions
		over all such source distributions.

An inseparable property of these Green's functions,
	which hitherto had little significance
		of its own,
is that \emph{all of the component sinusoids
	would have the same starting phase of zero},
as the phase factor $e^{i k s}$ in the Green's functions
	does not include
		a phase offset.
This is more general than
	the zero starting phase of differential beat waves,
and more particularly implies that
	the phase offsets of the Green's function solutions
		at a distance $x$ from the point source 
would be equal to $kx$,
	and hence proportional to frequency,
		since $\omega = k c$.
As a result,
the slope of the phase spectrum
		would be proportional to $x$,
as illustrated in
	Fig.~\ref{f:phasegrad},
which leads to
	the next result.

\begin{lemma}[Phase spectral slope] \label{t:phaseslope} 
	Given a travelling waveform $f(x)$,
	the distance to its point of origin is well indicated by
		the slope of its phase spectrum
	at large distances from the origin.
\end{lemma}

\begin{proof}
	By definition,
	the Fourier decomposition of $f$
		and its inverse are
	\begin{equation} \label{e:fourier}
		F(k)
		=
		\int_{-\infty}^{\infty}
			f(x)
			\,
			e^{- i k x}
			\,
			dx
	\quad
	\text{and}
	\quad
		f(x)
		=
			\frac{1}{2\pi}
		\int_{-\infty}^{\infty}
			F(k)
			\,
			e^{i k x}
			\,
			dk
	\quad
		.
	\end{equation}
	The inverse transform more particularly expresses
		the travelling waveform
	as the sum of
		its spectral components
		$F(k) \, e^{i k x}$.
	In signal processing theory,
	the phase of an arbitrary complex-valued signal
		$
		z(t)
		\equiv
			r(t)e^{i\phi(t)}
		$
	is defined as $
		\phi(t)
		=
			\arg(z(t))
		\equiv
			-i \log(z(t))
		$
	restricted to
		$(-\pi, \pi]$.
	Hence,
	the slope of the phase spectrum is
		the gradient
	\begin{equation} \label{e:phaseslope}
		\frac{\partial \phi}{\partial k}
		\equiv
			\frac{\partial}{\partial k}
				[
				i^{-1}
				\log ( F \, e^{i k x} )
				]
		=
			i^{-1}
			\frac{\partial}{\partial k}
				[
				\log  F
				+
				i k x
				]
		=
			x
		+
			i^{-1}
			\frac{\partial \log F}{\partial k}
	\quad
		.
	\end{equation}
	The continuity of the spectrum,
	which is assured
		only for travelling waves,
	is necessary for
		defining the gradient.
	The derivative last term in
		equation (\ref{e:phaseslope})
	comprises
		the starting phase variations,
	and is
		therefore constant.
	As a result,
	its relative contribution would diminish
		with distance,
	\ie,
	\begin{equation}
		\frac{ \partial \phi / \partial k } { x }
		=
			1
		+
			\frac{\partial \log F / \partial k}{ix}
		\longrightarrow
			1
	\quad
		\text{as}
	\quad
		x
		\longrightarrow
			\infty
	\quad
		,
	\end{equation}
	since $F$ is independent of $x$.
\end{proof} 


Lemma \ref{t:phaseslope} is a more general result
	as we would not be dependent on measuring
		a differential beat phase,
but measuring the slope still entails
	accurate measurement and correlation of the phase
		over at least two frequencies
	in the received spectrum.
Remarkably,
we can exploit
	the phase spectral slope
without 
	any phase measurements,
so as to efficiently
	obtain the distances of the sources,
or
	separate their signals,
as will be explained in
	Section \ref{s:shift}.

Though novel to
	most physicists and optical engineers
and
	never successfully exploited before,
this roll-off of phase with distance
	is not totally unfamiliar to signals engineers%
	\footnote{
	Dr.~John A.~Kosinski,
		US Army RDECOM CERDEC I2WD,
	has privately mentioned it
		as a long held intuition.
	}. 
The closest notion in optics is
	the coherence length,
defined as
	$
	L_c
	= 
		\lambda^2 / \eta \, \Delta \lambda
	$,
where
	$\Delta \lambda$ is
		the half-power ($3~\deci\bel$) bandwidth
	around the wavelength of interest
		$\lambda$,
and
	$\eta$ is the refractive index of the medium.
Since even the most stable lasers are limited to
	coherence lengths of a few metres,
the linespreads $\Delta \lambda$
	are known to be nonzero,
and would correspond to
	the spectral continuity of individual wavepackets
		as illustrated in Fig.~\ref{f:fcont}.
In interferometry and holography,
the term coherence refers to
	phase correlation over incremental distances
	(see \cite[Chapter X]{BornWolf}).
Likewise,
Doppler theory involves counting of wavefronts
	only of individual wavelengths,
and is orthogonal to
	the phase variation across wavelengths
		considered here.
This will become especially clear
	in the next section.



\Section{shift}{Frequency shifts by Fourier switching} 


\begin{wrapfigure}[10]{r}{70mm}
	\centering
	\psfig{file=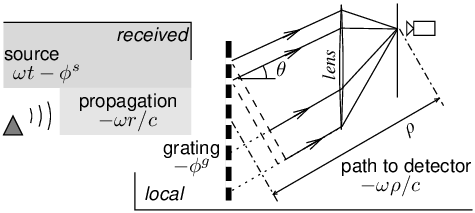, width=70mm}
	\caption{Components of phase}
	\label{f:phaseterms}
\end{wrapfigure}
In general,
a gradient $\partial \psi / \partial y$
	of a physical property $\psi$
transforms into
	a rate of change $d\psi/dt$,
if its domain is traversed
	at a steady speed $dy/dt$,
since
	$
	(\partial \psi / \partial y)(dy/dt)
	=
		d\psi/dt
	$.
This is especially useful in cases where
	$d\psi/dt$ is easier to measure
than
	$\psi$ or $\partial \psi/\partial y$.

The phase spectrum slope
	$\partial \phi / \partial k$
	(equation \ref{e:phaseslope})
is
	difficult to measure,
but a rate of change of phase,
	$d\phi/dt$,
is generally easy,
	since a rate of change of phase is
		either a frequency or a frequency shift.
The required domain traversal is
	a continuous variation of
		the receiver's frequency or wavelength scale
	at the rate $dk/dt$.

Fig.~\ref{f:phaseterms} depicts
	the various contributions to
		the instantaneous phase
at the photodetector in
	an instrument equipped
		with a diffraction grating and a lens
	for discriminating
		frequencies or wavelengths.
The total instantaneous detector phase $\phi''$ includes
	temporal variation $\omega t$
		starting at the source
			with a static phase offset $-\phi^{s}$;
	path delay from the source
		to the grating $- \omega r / c$;
	an additional phase change $- \phi^{g}$
		due to transmission by the grating;
and
	the path delay $-\omega \rho/c$
		from the grating to the detector, 
$\rho$ denoting
	the grating-detector path length.
The total instantaneous phase $\phi''$
	at the detector is thus
\begin{equation} \label{e:staticdetphase}
	\phi''
	=
		\omega t
		- \phi^{s}
		- \omega r / c
		- \phi^{g}
		- \omega \rho / c
\quad
	,
\end{equation}
so that the frequency actually detected would be
\begin{equation} \label{e:staticdetfreq}
	\frac{d \phi''}{dt}
	=
	\frac{d}{dt}
		(
		\omega t
		- \phi^{s}
		- \omega r / c
		- \phi^{g}
		- \omega \rho / c
		)
	=
		\omega
		(
		1
		- v/c
		)
\quad
	,
\end{equation}
where $\omega$ is the angular frequency
	selected by the diffraction angle $\theta$,
and
	$v \equiv dr/dt$ is
		the relative velocity from the source.
The second term in the final result
	accounts for the Doppler effect.
The static offset $\phi^{s}$ at the source,
	the grating shift $\phi^{g}$,
and
	the instrument path delay term $\omega \rho/c$
		do not have any effect
	on the detected frequency.
The signs were chosen
so that
	the wave phase could be written as
	$
	\phi(r,t)
	=
		\omega (t - r / c)
	$.
This is the ordinary case taken for granted
	in current spectroscopy,
and it would be straightforward to add
	relativistic corrections.

The frequency scale can be varied in
	one of two ways
to exploit
	the phase spectrum slope $\partial \phi/\partial k$.
We could move the detector
	over the focal plane of the lens,
so as to select
	successive frequencies $\omega'$
		at a rate $dk'/dt$,
as suggested by
	the detector scope and scale markings depicted
		at the right of Fig.~\ref{f:phasegrad}.
Or we could vary
	the grating intervals
so as to cause
	the frequency $\omega$ arriving
		at a fixed diffraction angle $\theta$
	to vary at the rate $dk/dt$.
In either case,
\emph{the detector would be presented
	a continuous sequence of instantaneous values
		from a succession of Fourier (sinusoidal) components
	of the travelling waveform,
in place of
	a single Fourier component},
so that
the phase at the detector would be accelerated
	relative to the static case
		in equation (\ref{e:staticdetfreq}),
with the following result.


\begin{theorem}[Phase acceleration] \label{t:shifts} 
	To an instrument with frequency scale changing 
		at a normalized rate $\beta$,
	the spectra of
		static sources
	will appear scaled in proportion to
		the source distances $r$
	as
	$
	\omega''
	\simeq
		\omega
		(
		1
		+ \beta r / c
		)
	$
	or, equivalently, at scale factors
	$
	z(r)
	\equiv
		\delta \omega / \omega
	\equiv
		(\omega'' - \omega) / \omega
	\simeq
		+ \beta r / c
	$
		at $r \gg \lambda$.
\end{theorem}

\begin{proof}
Assuming,
	as in Lemma \ref{t:sourceinfo},
that the amplitude varies slowly
	in the arriving travelling wave spectrum,
the variation of phase
	$\phi''$
	at the detector
will be primarily due to
	the phases of the Fourier components
		and other phase contributions
	already noted
		in equation (\ref{e:staticdetphase}),
and not so much from
	the amplitude differences between
		the Fourier components.
Then, to a first order,
the rate of change of phase at the detector
	would be
\begin{equation} \label{e:phaserates}
	\frac{d \phi''}{dt}
	=
		\left[
			\frac{d \phi}{dt}
		-
			\frac{d \phi^{s}}{dt}
		-
			\frac{d (\omega r / c)}{dt}
		\right]
		-
		\left[
			\frac{d \phi^{g}}{dt}
		+
			\frac{d (\omega \rho / c)}{dt}
		\right]
\end{equation}
where the first three terms on the right refer to
	the rate of change of phase
		entering the grating,
	and
	the remaining two,
		the rate of change
	due to the grating itself.
We have rewritten $\omega t$ as $d\phi/dt$
to allow for the fact that
	$\omega$,
representing
	the instantaneously selected Fourier component,
		would be itself changing.

We shall represent
	the instantaneous selection by a single prime,
	as $k'$.
The first term on the right
	in equation (\ref{e:phaserates})
then leads to two parts,
	the intrinsic rate of change
		due to the original angular frequency $\omega$,
and a part reflecting
	the changing wavelengths
		at the same diffraction angle $\theta$,
as
\begin{equation} \label{e:srcderiv}
	\frac{d \phi}{dt}
	=
		\frac{\partial \phi}{\partial t}
	+
		\frac{\partial \phi}{\partial k'}
		\frac{dk'}{dt}
	\equiv
		\omega
	+
		\dot{k}'
		\,
		\frac{\partial \phi}{\partial k'}
	=
		\omega
\quad
	,
\end{equation}
since diffraction \perse
	does not contribute to phase
nor, hence, to
	its rate of change;
dispersion by the grating will be accounted for
	separately below by terms in $\phi^{g}$.
The second term on the right
	in equation (\ref{e:phaserates})
expands similarly,
except 
	$
	\partial \phi^{s} / \partial t
	\equiv
		0
	$
as $\phi^{s}$
	are constants with respect to time.
However,
	$
	\dot{k}' \, \partial \phi^{s}/\partial k'
	$
survives because
	$\phi^{s}$ would likely vary across wavelengths,
		and hence across successive values of $k'$.
Modulation at the source
would contribute to
	variation of $\phi^{s}$ across wavelengths,
and would be hence contained
	in this term.
For constant wave speed $c$,
the remaining terms similarly expand to
\begin{equation} \label{e:miscderiv}
\begin{split}
	\frac{d (\omega r / c)}{dt}
&
	=
		\dot{r}
		\,
		\frac{\partial (k r)}{\partial r}
	+
		\dot{k}'
		\,
		\frac{\partial (k r)}{\partial k'}
	=
		k \dot{r}
	+
		\dot{k}' r
	\quad
	,
\\
	\frac{d \phi^{g}}{dt}
&
	=
		\frac{\partial \phi^{g}}{\partial t}
	+
		\dot{k}'
		\,
		\frac{\partial \phi^{g}}{\partial k'}
	\quad
	\text{respectively, and}
\\
	\frac{d (\omega \rho / c)}{dt}
&
	=
		\dot{k}'
		\,
		\frac{\partial (k \rho)}{\partial k'}
	=
		\dot{k}' \rho
	\quad
	,
\end{split}
\end{equation}
noting that
	$k \equiv \omega / c$
		generally,
and
	$\dot{\rho} = 0$
		for a static lens assembly.
Also,
$\partial \phi^{g}/\partial t$ cannot vanish
	while the grating is being varied,
but as remarked above,
we could move the photodetector instead
	and keep the grating fixed,
in which case,
	we would indeed have
	$\partial \phi^{g}/\partial t = 0$.
In either case,
equation (\ref{e:phaserates})
	leads to
\begin{equation} \label{e:allterms}
\begin{split}
	\frac{d\phi''}{dt}
	\equiv
		\omega''
	=
		\omega
	-
		k \dot{r}
	-
		\dot{k'}
		(
			r
		+
			\rho
		+
			\phi^{s}_{,k'}
		+
			\phi_{,k'}^{g}
		)
\quad
	( _{,k'} \equiv \partial/\partial k' )
\quad
	.
\end{split}
\end{equation}
Absorbing $\rho$ into $r$,
	replacing
		$\dot{r}$ with $v$,
and noting that
	$
	k' 
	\equiv
	k
	$
because
	the output and input wave vectors at the grating
		would be identical at each instant,
we get
\begin{equation} \label{e:z}
\begin{split}
	\omega''
	\equiv
&~
		\dot{\phi}''
	=
	~
		\omega
		(
		1
		- v / c
		+ [ r + ( \phi_{,k}^{s} + \phi_{,k}^{g} ) ] \beta / c
		)
	~
	\simeq
	~
		\omega
		(
		1
		- v / c
		+ \beta r / c
		)
\quad
	,
\\
\text{and~}
	z(r)
	\equiv
&
	~
		\delta \omega / \omega
	\equiv
	~
		(\omega'' - \omega) / \omega
	=
	~
		- v / c
		+ [ r + ( \phi_{,k}^{s} + \phi_{,k}^{g} ) ] \beta / c
	~
	\simeq
	~
		- v / c
		+ \beta r / c
\quad
	.
\end{split}
\end{equation}
The asymptotic form generally holds
	for $r \gg \lambda$,
since
	the source phase modulation
and 
	grating delay variation
would be unlikely to exceed
	a few wavelengths.
To prove that the spectrum would be indeed scaled by
	the factor
	$
		(
		1
		- v / c
		+ \beta r / c
		)
	\equiv
		\Delta
	$,
it is convenient to use
	the bra-ket notation of quantum mechanics.
Fourier decomposition in general is
	governed by the orthogonality condition,
	$
	\iprod{\omega''}{\omega,r}
	\equiv
		\int
		e^{-i \omega'' t}
		\,
		e^{-i (k r - \omega t) }
		\, dt
	=
		e^{-i k r}
		\delta(\omega'' - \omega)
	$.
As $t$ literally denotes time
	\emph{measured by the Fourier analyzer clock},
the orthogonality should allow for
	a variability in the ``ticking'' of $t$,
and thus
	more generally written as
	$
	\iprod{\omega''}{\omega,r}
	\equiv
		\int
			e^{-i \omega'' t}
			\,
			e^{i (d \phi'' / dt) (t - r/c) }
		\, dt
	$.
Identifying $d\phi''/dt$ as
	the phase at the detector,
and applying
	the asymptotic forms in equations (\ref{e:z}),
we get
\begin{equation} \label{e:rorth}
\begin{split}
	\iprod{\omega'',\beta}{\omega,r}
&
	\equiv
		\int
		e^{-i \omega'' t} \,
		e^{i (d \phi'' / dt) (t - r/c) }
		\, dt
	=
		\int
		e^{-i \omega'' t} \,
		e^{i \omega (1 - v/c + \beta r / c) (t - r/c)}
		\, dt
\\&
	=
		e^{-i (k \Delta) r}
		\delta(\omega'' - \omega \Delta)
	,
\quad
	\Delta
	\equiv
		(1 - v / c + \beta r / c)
\quad
	.
\end{split}
\end{equation}
For an arriving waveform
	$
	\ket{f,r}
	\equiv
		f(t - r/c)
	$,
the notation
	$\iprod{\omega}{f} \equiv F(\omega)$
yields
	$
	\iprod{\omega''}{f,r}
	\equiv F(\omega'',r)
		= e^{-ikr} F(\omega'')
	$.

\begin{wrapfigure}[6]{l}{70mm}
	\centering
	\psfig{file=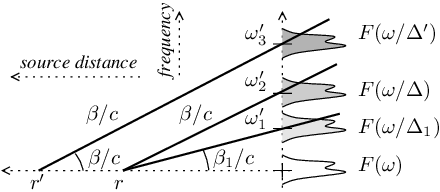, width=70mm}
	\caption{Time domain parallax}
	\label{f:parallax}
\end{wrapfigure}
In the presence of phase acceleration,
	$
	\iprod{\omega''}{f,r}
	$
changes to
\begin{equation} \label{e:zwaves}
\begin{split}
&
	\iprod{\omega'',\beta}{f,r}
	\equiv
	\int_t
	\int_\omega
		\iprod{\omega'', \beta}{t}
		\,
		dt
		\,
			\iprod{t}{\omega}
			\,
			d\omega
			\,
			\iprod{\omega}{f,r}
\\&
	=
	\int_{\omega}
	\int_{t}
		e^{-i \omega'' t}
		\,
		dt
		\,
		e^{i (d\phi''/dt) (t - r/c)}
		F(\omega)
		\,
		d\omega
\\&
	=
	\int_\omega
		e^{-i k'' r}
		\delta(\omega'' - \omega \Delta)
		F(\omega)
		\,
		d\omega
	=
		e^{-i k'' r}
		F(\omega''/\Delta)
\quad
	,
\end{split}
\end{equation}
using equation (\ref{e:rorth}) to reduce
	the time integral.
Equation (\ref{e:zwaves}) proves that
	the scaling would be real, uniform and proportional
		to the source distance $r$,
and therefore
	also distinct from
		the changing scale of the instrument.
It is straightforward to prove,
by applying the superposition principle
	to the travelling waveforms,
that equations (\ref{e:staticdetphase})-(\ref{e:z})
would also hold for
	a multitude of
		travelling waveforms $f_j(x)$
	arriving simultaneously from
		sources at different distances $r_j$,
and would yield
	the scale factors
	$
	\Delta_j
	\simeq
		(
		1
		- v_j / c
		+ \beta r_j / c
		)
	$,
	for the respective waveforms,
		for the same common $\beta$,
so that
	the spectra would be separated in frequency
		according to the source distances,
	as illustrated in Fig.~\ref{f:parallax}.

For instance,
for an astronomical source at
	$
	r / c
	\sim
		10
		~\giga\yyear
		\footnote{
			Giga year
				$\equiv 10^9$ years.
			Geological literature also uses 
				Ma and Ga
			to denote
				mega- and giga-annum,
			respectively.
		}
	\equiv
		3.156 \times 10^{16}
		~\second
	$,
	$
	\beta
	\approx
		3.17 \times 10^{-17}
		~\reciprocal{\second}
	$
would suffice for
	$z = 1$,
	\ie, doubling in frequency.
This is so small that
	the total change of the grating intervals
	$
	\zeta(\Delta t)
	=
		\int_{\Delta t}
			\beta
			\,
			d\tau
	$
would amount to just
	$1.13 \times 10^{-14}$,
or a little over a cycle at visible wavelengths,
	for observation times of $\Delta t = 1~\hour$,
and only $10^{-10}$
	hypothetically for a full year.
Larger $\beta$ would be needed
	in most practical applications,
but the shifts $\beta r/c$ 
	would remain larger than, and distinct from,
		$\zeta$ over
			any individual Fourier integration window.
\end{proof} 


\Section{comm}{Distance multiplexing, jamming immunity and passive ranging} 

\begin{wrapfigure}[09]{r}{55mm}
	\centering
	\psfig{file=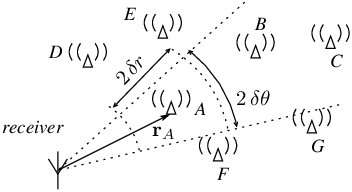, width=55mm}
	\caption{Multiplexing by distance}
	\label{f:ddm}
\end{wrapfigure}

The principal problem of communication engineering,
	as remarked in the Introduction,
is facilitating
	independent communication between
		multiple pairs of locations.
In terrestrial communication,
	these end points generally lie in
		the two-dimensional space
	of the earth's surface,
		as illustrated in Fig.~\ref{f:ddm}.
The separation in frequency by source distance
	allows separation of the respective signals
using the physics of space
	instead of modulation or content.
For example,
	in Fig.~\ref{f:parallax},
the same spectrum $F(\omega)$
	could be scaled to
	$
	F(\omega/\Delta_1)
	$,
where
	$
	\Delta
	\equiv
		1 + \beta_1 r/c
	$,
or to
	$
	F(\omega/\Delta)
	$,
	$
	\Delta
	\equiv
		1 + \beta r/c
	$
if it came from a source at distance $r$,
	depending on whether
		the rate of change was $\beta$ or $\beta_1 < \beta$,
assuming the source and the receiver
	are relatively static.
For the same $\beta$,
the same spectrum would get scaled to
	$
	F(\omega/\Delta')
	$,
where
	$
	\Delta'
	\equiv
		1 + \beta r'/c
	$
if it came from another (static) source
	at distance $r'$.
Then, with a band-pass filter,
	we may isolate either
	$F(\omega/\Delta)$
or
	$F(\omega/\Delta')$,
and recover
	the original signal from
		the corresponding source
by scaling
	the filtered spectrum back to
		the original frequency band.
A formalism for treating this is given next.


\begin{corollary}[Cascading phase acceleration] \label{t:opgroup} 
	Phase acceleration can be cascaded, and
	the product operator representing
		cascading of phase accelerations
	forms a commutative group
		for small $\beta$.
\end{corollary}

\begin{proof}
	We may define
		a \emph{phase acceleration operator} $H(\beta)$
	by setting
		$
		\bra{\omega'',\beta}
		= 
			\bra{\omega''} H(\beta)
		$,
	so that
		$
		\iprod{\omega'',\beta}{f,r}
		\equiv
			\bra{\omega''} H(\beta)
			\ket{f,r}
		$.
	We would then have $H(0) = 1$,
		defining the identity element,
	since
		$
		\bra{\omega''} H(0) \ket{f,r}
		\equiv
			\iprod{\omega''}{f,r}
		\equiv
			e^{-ikr} F(\omega'')
		$,
	which preserves
		the received spectrum.
	The inverse is similarly given by
		$H^{-1}(\beta) = H(-\beta)$.
	From the definition
		$k' \equiv 2 \pi / \lambda'$
	which relates $k'$ to
		the instantaneously selected wavelength
			$\lambda'$,
	we have
		$
		k'^{-1} dk'/dt
		\equiv
			(2 \pi / \lambda')^{-1}
			d (2 \pi / \lambda')/dt
		=
			\lambda'
			. (- \lambda'^{-2} d\lambda'/dt)
		=
			- \lambda'^{-1} d\lambda'/dt
		$,
	\ie, \emph{a negative $\beta$ implies
		upward scaling of wavelengths
			instead of frequencies}.
	In an all-optical system
		using diffractive elements for both
			phase acceleration and filtering,
	we would thus need to use
		negative $\beta$
	for the time domain parallax
		and signal separation.

	Since the waveform arriving at the detector in
		Figs.~\ref{f:phasegrad} and \ref{f:phaseterms}
	already presents
		a scaled wavelength or frequency,
	it can be subjected to
		a further phase acceleration
			using a second grating
		with continuously changing intervals,
	to which equations (\ref{e:allterms})
		would be applicable once again.
	This provides a product operation for the $H$ operators,
	and the product scale factor
		corresponding to
		$
		H(\beta_1) H(\beta_2)
		$
	would be
		$
			(1 + \beta_1 r/c)(1 + \beta_2 r/c)
		=
			(1
			+ [\beta_1 + \beta_2] r/c
			+ \beta_1 \beta_2 \, r^2/c^2
			)
		\approx
			1 + (\beta_1 + \beta_2) \, r/c
		$
	for
		$\beta_1, \beta_2 \ll c/r$.
	This yields
		$
		H(a \beta_1) H(b \beta_2)
		=
			H(b \beta_2) H(a \beta_1)
		\approx
			H(a \beta_1 + b \beta_2)
		$,
	so that the product is commutative
		and linear for small $\beta$,
	and makes
		the set of $H(\beta)$ operators a group.
\end{proof} 


The signal selection and separation process 
	would be described by
the product operator
	$
	H^{-1}(\beta)
	\,
	\widetilde{G}_m
	\,
	H(\beta)
	\,
	G_b
	$,
where
	$G_b$ is
	the band-pass prefilter admitting
		the unscaled (unaccelerated) received spectrum 
		$F(\omega)$ as in Fig.~\ref{f:parallax};
	$H(\beta)$ is
	the first phase acceleration employed to
		spread the received prefiltered signal
		to
			$F(\omega/\Delta)$
		and
			$F(\omega/\Delta')$;
	$\widetilde{G}_m$ is
	a band-pass ``source selection'' filter
		for selecting either
			$F(\omega/\Delta)$
		or
			$F(\omega/\Delta')$;
and
	$H^{-1}(\beta) \equiv H(-\beta)$ is
	the ``phase deceleration''
		for returning the selected scaled spectrum
			back to its original frequency band.

Similar notions of
	\emph{space division multiple access} (SDMA)
and of
	\emph{space division multiplexing} (SDM)
characterize
	multibeam antennas
	on satellites and
	in orthogonal frequency division multiplexing (OFDM).
These existing notions do not include
	multiplexing by distance,
		however.
The connotation of ``division''
in these terms is also at slight variance from
	the traditional notions of time, frequency or code division
		multiplexing or multiple access,
	in which
		the channel capacity itself becomes divided,
whereas
	the distance division capability,
	like the division of the solid angle
		in satellite SDMA,
\emph{multiplies}
	the overall channel capacity,
as 
	the full bandwidth can be
		reused by each source.
The multiplication factor is quantified
	as follows.


\begin{corollary}[Distance multiplexing] \label{t:ddm} 
	Any receiver using phase acceleration can resolve
		indefinitely many band-limited transmissions
	at increments
		asymptotically proportional to the distance.
\end{corollary}

\begin{proof}
	Assuming a centre frequency $f_c$
	and
		nominal signal bandwidth $2W$
			including guard bands,
	a pair of
	sources at distances $r$ and $r' > r$ (Fig.~\ref{f:parallax})
		would be just resolved
	when the higher end frequency $f_c + W$
		from the source at $r$
	scales to just below
		the lower end frequency $f_c - W$
			from the source at $r'$,
	\ie,
		$
		(f_c + W)(1 + \beta r/c) = (f_c - W)(1 + \beta r'/c)
		$.
	Rearranging and substituting
		$r_n$ for $r$
	and
		$r_{n+1}$ for $r'$,
	yields
		the infinite recursion
	\begin{equation} \label{e:srcstep}
	\begin{split}
		\beta r_{n+1} / c
		=
			\gamma
		+
			(1 + \gamma)
			\,
			\beta r_n / c
	\quad
	&
	\text{, where}
	~
		\gamma
		\equiv
			2W /(f_c - W)
	\text{~and~}
		n = 1, 2, \dots
		;
	\\
	\text{whence}
	\quad
		\delta r_{n}
		\equiv
			r_{n+1} - r_n
		=
			\gamma \, (r_n + c/\beta)
	\quad
	&
	\text{, so that}
	\quad
		\frac{\delta r}{r}
		\simeq
			\gamma
		\simeq
			\frac{2W}{f_c}
	\quad
	\text{for~}
		f_c \gg W
	\quad
		.
	\end{split}
	\end{equation}
The result is
	a distance resolution of $\pm \delta r$
		proportional to $r$,
	similar to parallax.
\end{proof} 


If combined with a directional antenna for  
	azimuth and elevation resolutions of
		$\pm \delta \theta$ 
	and
		$\pm \delta \psi$,
	respectively,
this would enable the receiver
to selectively listen to a transmitter at
	any desired coordinates
		$(r, \theta, \psi)$
	relative to the receiver,
regardless of all other transmissions
	on the same frequencies,
		friendly or otherwise,
as follows.


\begin{corollary}[Noise and jamming immunity] \label{t:jamming} 
	Interference or noise originating at sufficient distance
		from a source of a band-limited signal
	can be suppressed
		independently of signal content,
	using phase acceleration.
\end{corollary}

\begin{proof}
	Rejection of out-of-band received interference and noise
		depends on the performance of
			the prefilter $G_b$.
	Rejection of in-band interference and noise
		from noncolocated sources depends mainly on
			the selection filter $\widetilde{G}_m$.
	We would also expect noise and distortion
		due to nonuniformity
			and thermal or mechanical fluctuations
		in the diffraction grating or other means
			used for the phase acceleration operations
				$H$ and $H^{-1}$.
	Fluctuations in $\beta$ would introduce interference
		as they would cause the signals from
			neighbouring sources to overlap.
	The broad result holds for
		interference and noise sources
	outside the selected distance interval
		$r \pm \delta r$.
\end{proof} 


Lastly,
the shift $\beta r/c$ itself is
	a linear instant measure of source distance,
available without round-trip timing
	or phase correlation with
		another illuminating source,
as required in
	both active and passive radars today.
For visual indication,
it would suffice to apply
	a notch prefilter to the received spectrum,
and to display
	a precomputed distance scale alongside the frequency
		(or wavelength) axis
	with no signal correlation at all.



\Section{other}{Wavelength transformation and synthesis} 

As $\beta$ is
	a rate of change of
		receiver frequency scale
and both
	$\beta$ and the range of variation
	are limited only by
		the state of technology,
phase acceleration enables
	continuous wavelength scaling
over
	any range
by
	any factor,
well beyond the capability of
	bandgap and nonlinear materials.
The scaling could be additionally used
	for accurately synthesizing radiation
		of any wavelength
	with desired
		polarization, coherence or modulation,
by applying it to
	an available source already capable of providing
		the latter properties.
As both $\beta$ and the source distance $r$
	would be usable for fine control,
accuracy and continuous tunability of
	the output wavelength are assured.

\begin{wrapfigure}[11]{r}{85mm}
	\centering
	\psfig{file=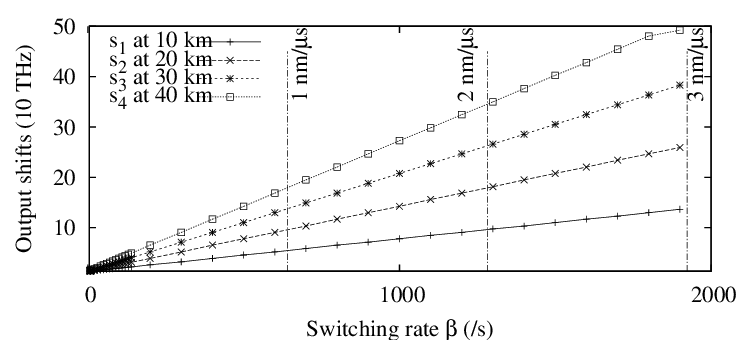, width=85mm}
	\caption{Computed shifts for emerging fast FBGs}
	\label{f:separation}
\end{wrapfigure}
An inherent limitation lies in the fact that
	phase acceleration expands or contracts
		successive segments of the input waveform.
If the wavelength is contracted,
	the output would contain gaps
		between sweeps of
			the frequency scale variation;
if expanded,
	portions of the source waveform
		would be correspondingly skipped.


\Section{veri}{Realizability and partial verification} 

For device and terrestrial communication applications,
	we require $z = O(1)$
at distance scales of
	under $1~\metre$ to about $10^4~\metre$,
which calls for
	normalized rates of change
	$
	\beta
	\equiv 
		z c / r
	\sim
		10^{8}
		\text{~to~}
		10^{4}
		~\reciprocal{\second}
	$.
Even for $z \sim 10^{-6}$
	at $r \approx 1~\metre$,
we would need
	$
	\beta
	\sim
		300
		~\reciprocal{\second}
	$.
This may still seem large,
but the continuous variation is only needed
	in repetitive sweeps
that can be as short as
	$
	1
		~\micro\second
	$.
The total variation would be
	$
	\zeta(1~\micro\second)
	\equiv
		\exp(300 ~\reciprocal\second \times 1~\micro\second)
		- 1
	\approx
		0.0003
	$.
This is in the range of
	piezoelectric or magnetostrictive elements,
on which
	a reflective plastic grating could perhaps be bonded,
but the speed 
	would be in milliseconds.

\begin{wrapfigure}[8]{l}{75mm}
	\centering
	\psfig{file=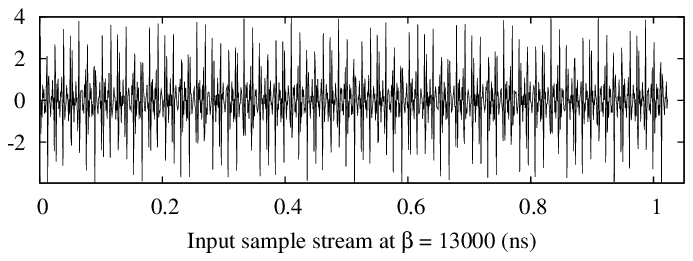, width=80mm, height=26mm}
	\caption{Distortion from variable sampling}
	\label{f:distortion}
\end{wrapfigure}
Current fast-tunable fibre Bragg gratings (FBGs)
	that use mechanical stress
are rated at switching speeds of
	only a few $\nano\metre$ in wavelength per $\milli\second$
		at around $1.5~\micro\metre$,
for which
	$
	|\beta|
	\sim
		(1.5~\micro\metre)^{-1}
		\times
		1
		~\nano\metre
		~\reciprocal{\milli\second}
	\approx
		0.7
		~\reciprocal{\second}
	$.
At these rates,
a barely observable shift
	of $z = 10^{-6}$
would need
	$
	r
	\equiv c z / \beta
	\approx
		500
		~\kilo\metre
	$,
which makes the sun
	the nearest available test source!
This would also explain
	why phase acceleration and the associated shifts
		remain unnoticed.

Only the next generation devices,
expected to attain
	Fourier switching speeds in the order of
		$\nano\metre$ wavelength
	in
		$\micro\second$ or $\nano\second$%
	\footnote{
	See, for example,
		the Sabeus and Optonet abstracts at
	\texttt{http://www.dodsbir.net/selections/sttr1\_05.htm}.
	}, 
would be able to produce measurable $z$
	within a laboratory.

Fig.~\ref{f:separation} illustrates
spectral scaling and signal separation
	over $10~\kilo\metre$ fibre spools
	between
		four $1.54~\micro\metre$ ($\equiv 130~\tera\hertz$)
	sources,
assuming a core index of
	$\eta \approx 1.47$,
simulated in
	an equivalent digital signal processing (DSP)
		realization of phase acceleration
	using variable sampling
		\cite{Prasad2006a,Prasad2005a}.
A Java applet developed for this test incorporates both
	deterministic and statistical
		simulations of linespreads
	in accordance with Lemma \ref{t:cont}.
The sampling rate variation $\beta$ is
	independent of the source distances,
which are input only to
	the path phase computation,
yet the unmodified open source Java FFT
	applied to the sample stream
		reveals the shifts,
	as a partial test of the theory.
The time domain signal is distorted
	by the sampling rate variation,
but is still periodic,
	as shown in Fig.~\ref{f:distortion}.


\Section{broad}{Broader implications} 


The principle of radiation quantization and,
more particularly,
	Planck's quantization rule
		$E = \hbar \omega$
have made it too easy to view
	electromagnetic radiation as comprising
		monochromatic quanta.
This corpuscular perception should be probably blamed for
	the inadequate treatment of diffraction
		in both particle physics and astrophysics,
	to be pointed out below.
Although the second quantization formalism
	is in fact based on standing wave modes,
it is commonly overlooked that
	travelling quanta cannot be monochromatic,
		and therefore cannot be Planckian!
Conversely,
Planck quanta are necessarily\
	merely detector state transitions,
under no fundamental obligation to exactly correspond to
	source state transitions
		responsible for the radiation.
The Fourier integration involved in
	any form of spectroscopy or frequency or wavelength selection
is
	an inherently classical \emph{macroscopic} process,
and also
	under no obligation to preserve
		the original quanta.
Although multi-quanta interactions are well known,
the possibility of \emph{reconstituting} photons
	by recombining
		multiple original Fourier components is new,
and introduces a more general notion of
	coherent integral transformations of light.
More significant implications of
	the present theoretic result
concern astrophysics,
	as follows.


\begin{principle}[Hubble uncertainty] \label{t:uncert} 
	Real spectrometers are subject to a fundamental uncertainty
		affecting distant spectra,
	given by
		$
		\Delta \beta \, \Delta T_E
		\approx
				1
		$,
	where
		$\Delta T_E$ is the time constant of
			component variations
	and
		$
		\Delta \beta
		\equiv
			\beta
		$
	is the resulting scale uncertainty
		affecting remote spectra
			according to equations (\ref{e:z}).
	Since creep cannot be totally eliminated,
		taking $
		t_\sun
		\equiv
			4.9~\giga\yyear
		=
			1.55 \times 10^{17}
			~\second
		$, the age of the sun,
	as a conservative upper bound
		on the rigidity of our instruments,
	\ie,
		$
		\Delta T_E
		\le
			t_\sun
		$,
	yields a scale uncertainty of
		$
		\Delta \beta
		\ge
			t_\sun^{-1}
		=
			200
				~\kilo\metre
				~\reciprocal{\second}
				~\reciprocal{\mega\parsec}
		\approx
			2.7~H_0
		$,
	the Hubble constant.
\end{principle}

\emph{Rationale.}
	Creep is usually associated with
		high temperatures and stresses,
	but solid state physics provides
		no fundamental constraint prohibiting
			dislocations
		even at arbitrary low stresses.
	The residual creep rate would be governed by
		the dislocation probability
		$ p_d \equiv e^{-W_d / k_B T} $,
	where
		$k_B$ is the Boltzmann constant,
		$T$ is the operating temperature
	and	
		$W_d$ is the dislocation energy barrier
			typically about $1$-$2~\eV$,
	hence
		$
		p_d
		\approx
			10^{-11} \dots 10^{-21}
			~\reciprocal{\second}
		$.
	We should expect residual creep,
		under the compression of earth's own gravity
		(due to the gradient $\nabla \mathbf{g}$)
		and aided by the kneading effect of tides,
	to cause extremely slow continual shrinkage
		in instrument dimensions,
	hence a negative $\beta$, consistent with redshifts,
		of within a few orders of $p_d$.
	The Hubble redshifts are in fact of this order,
	since $
		H_0
		\approx
			73
				~\kilo\metre
				~\reciprocal{\second}
				~\reciprocal{\mega\parsec}
		=
			2.4 \times 10^{-18}
				~\reciprocal{\second}
		$.
$\Diamond$


Two further results reinforce this implication.
First, again from equations (\ref{e:z}),
	we may equate $v = \beta r / c$
to obtain
	the equivalent apparent velocity of remote sources.
With $\beta \sim H_0$,
	this would have the same form as
		the cosmological expansion.
Since the motion would be apparent and not real, however,
	it would be perfectly linear,
so that in the associated acceleration
	$
	\dot{v} 
	\equiv
		\beta \dot{r}/c
		+
		r \dot{\beta}/c
	=
		\beta^2 \, r/c
		+
		r \dot{\beta}/c
	$,
we should set
	$\dot{\beta} = 0$ identically,
since the acceleration refers to
	the optical path time (history) of the received light,
and not
	the instantaneous variation at the instrument. 
In terms of relativistic cosmology,
	this acceleration component of the uncorrected shifts
would be characterized by
	the deceleration coefficient
	$
	q
	\equiv
		(-1 + \dot{\beta}/\beta^2)
	=
		-1
	$.
The observed cosmological acceleration
	in fact corresponds to
	$
	-1 \pm 0.4
	$
	\cite{Riess1998}.
The next result is also unmatched by
	alternative cosmology theories.


\begin{theorem}[Spectrometer and Doppler time dilations]\label{t:td}
	Both spectrometric scale drift and Doppler shifts imply
		a commensurate, apparent time dilation
			in the received light.
\end{theorem}

\begin{proof}
	From equation (\ref{e:zwaves}),
		writing $\tau$ for receiver time
	and
		requiring
		$
		\iprod{\tau, -\beta}{\omega'', \beta}
			\equiv
			e^{i \omega''\tau}
		$
	leads to
	\begin{equation} \label{e:zevents}
	\begin{split}
		\iprod{\tau, -\beta}{f, r}
	&
		\equiv
		\int_{\omega''}
			\iprod{\tau, -\beta}{\omega'', \beta}
			\,
			d\omega''
			\,
			.
			\,
			e^{-i k'' r}
			F(\omega''/\Delta)
		=
		\int_{\omega''}
			e^{i \omega''\tau}
			\,
			d\omega''
			\,
			.
			\,
			e^{-i k'' r}
			F(\omega''/\Delta)
	\\
	[\text{substituting~} \omega' = \omega''/\Delta]
	\quad
	&
		=
		\int_{\omega'}
			e^{i \omega' \tau \Delta}
			e^{-i \omega' r \Delta / c}
			F(\omega')
			\,
			d\omega' \Delta
		\equiv
			\Delta
			.
			f([\tau - r / c]\Delta)
	\quad
		.
	\end{split}
	\end{equation}
	This matches
		the cosmological time dilation,
	which is remarkable because time dilation
		has been hitherto regarded
	as
		an exclusive consequence of relativity,
	so much so as to attribute
		the Doppler time dilation
	also to gravity
		\cite{Bolos2005}.

	All that equation (\ref{e:zevents}) represents
	is that
		a uniform scaling of frequencies
	means
		a reciprocal scaling of time.
	As the Doppler effect 
		uniform scales the whole spectrum, 
	repetition frequencies of pulses from a moving source
		would scale by the same amount.
	Time dilation looks exotic
	only because
		scattering and other matter interactions
			like the Wolf effect
			\cite{Wolf1987},
		do not uniformly act on
			the whole frequency axis,
	and in part because
		it is not significant
			in terrestrial applications.
	The next result concerns
		the related effect on luminosity.
\end{proof} 

\begin{principle}[Tolman's brightness law] \label{t:tolman} 
	The apparent brightness of distant sources
	should diminish as
		$
		\Delta^4
			\equiv
			(1 + \beta r / c)^4
		\equiv
			(1 + z)^4
		$
	with distance $r$ due to $\beta < 0$ from
		any systematic scale drift
			in our instruments.
\end{principle}

\emph{Rationale.}
	Tired light theory was proposed by Zwicky
		\cite{Zwicky1929}
		as a possible alternative to actual expansion
	almost immediately following
		Hubble's law
		\cite{Hubble1929}.
	Tolman proposed testing
		the reality of expansion
	by verifying whether
		the brightness of distant sources decreased,
			after removing the effects of dust extinction
				and peculiar velocities,
		with the redshifts as
			$(1 + z)^4$
	where
		one factor of $(1 + z)$ would be due to
			the photon energy reduction due to the redshift,
		a second $(1 + z)$ factor accounts for
			the decrease in the photon flux rate
				due to the expanding distance,
	and
		the remaining $(1 + z)^2$ accounts for
			larger apparent area at the time of emission
			(aberration),
	since in tired light theory,
		the brightness should decrease only due to redshift,
			as $(1 + z)$,
	as the distance expansion and area aberration factors
		would be absent
		\cite{Tolman1930,Tolman1934}.
	Actual tests yield exponents of $2.6$ or $3.4$
		depending on the frequency band
		\cite{Sandage2001,Lubin2001a,Lubin2001b,Lubin2001c},
	and
	the difference from the expected exponent $4$ matches
		the brightness variation
			predicted in stellar evolution
		(\cf \cite{Bruzual1993}).

	The shifts due to phase acceleration
		would be also equivalent to the Doppler effect
			of apparent radial motion,
	corresponding to
		$v/c \equiv - \beta r/c$
	or
		$v = - \beta r$
		in equations (\ref{e:z}),
	after eliminating
		peculiar motions.
	The aberration factor of
		$\Delta^2 \equiv (1 + z)^2$
		holds identically
	because when we interpret the shifts as Doppler,
		our resulting source models would be based on
			the Doppler-corrected past distances,
		just as assumed in Tolman's proposal,
	so that
		the observed brightnesses would have to be 
			similarly corrected for
				smaller starting areas
		for consistent physics.
	Time dilation between photons
		would seem to be given by the second $\Delta$
			in equation (\ref{e:zevents}),
	and it could be argued that
		the detector state transitions 
			representing the detected photons
		would be of lower energies as well.
	However,
	we can no longer assume 
		identity or even a 1:1 correspondence
	between
		the source emitted wavepackets
			and the detected photons
	since equation (\ref{e:zwaves}) signifies
		a \emph{reconstitution} of photon energies.
	All of the preceding theory is
		inherently \emph{classical}.
	The remaining $\Delta^2$ factor emerges \emph{classically}
		from equation (\ref{e:zevents}),
	as the power flux is given by
		$
		\int
			|\iprod{\tau, -\beta}{f, r}|^2
			\,
			d \tau
		\equiv
		\Delta^2
		\int
			|f([\tau - r / c]\Delta)|^2
			\,
			d \tau
		$.
$\Diamond$

\vspace{10pt}

Though Principles \ref{t:uncert} and \ref{t:tolman}
	primarily illustrate
the physics of phase acceleration
	and the associated frequency shifts,
specifically,
	the amplification of creep
and 
	the redistribution of power flow,
the implied challenge to current cosmology theory
	is unfortunately real.
The elementary conjugacy of
	time dilation to change of frequency scale,
		exposed in equation (\ref{e:zevents}),
is also not the only omission in
	current mainstream cosmology.

There has been no
	path-integral treatment of diffraction
for
	intra-galactic and cosmological space,
as required for
	a volumetric distribution of absorbing bodies,
which yields
	a propagation law of the form
	$
	I(r)
	\propto
		r^{-2}
		\,
		e^{-\sigma r}
	$,
similar to that
	for the ordinary attenuation of sound.
All of the diffractive considerations in both
	astronomy and particle physics
have been limited to
	Fresnel-Fraunhofer approximations
	(\cf \cite[pp.149-153]{Spitzer1978},
	\cite[pp.148-149]{Weinberg1995}),
which are inadequate for describing
	successive diffractions of wavefronts
that would lead to
	inaccessible trapped states of energy,
resembling neutrinos and contributing to
	both the radiation background and dark matter,
as the
	$
	r^{-1}
	\,
	e^{-\sigma r}
	$
	amplitude profile is
a Klein-Gordon eigenfunction for
	particles of effective mass
		$\hbar \sigma / c$.
Diffractive (and gravitational)
	$
	\sigma
	$
of just
	$
		0.05
		~\deci\bel
		~\reciprocal{\mega\parsec}
	\equiv
		10^{-24}
		~\deci\bel
		~\reciprocal{\metre}
	$
is all it takes to yield
	the same cutoff of the visible universe
		as the redshifts
	(\cf \cite{Wesson1987,Wesson1991})
and diffractive loss of metallic lines
	due to \emph{off-ray} proximal matter
is also unaccounted for
	in the treatment of primitive galaxies
	(\cf \cite{Elmegreen2005,Bournaud2007}).
The third key piece of the standard model,
	primordial nucleosynthesis,
again represents
	a predisposed model view,
since the observed nuclear abundances,
	such as of helium
	\cite{PenziasNobel1978}
	or lithium
	\cite{Korn2006},
	which are inconsistent with galactic production
		in the finitely old universe
			of current cosmology
		and call for
			a homogeneous large scale cooling,
would be equally and more simply explained by
	accumulation in an infinitely old universe
\emph{without} requiring
	large scale cooling or inflation theory.

Closer to home,
the measured lunar recession of
	$
	3.82
	\pm 0.07
		~\centi\metre
		~\reciprocal{\yyear}
	$
	\cite{Lunar1994}
exceeds
	a na\"ive linear Hubble's law expansion
		on short scale%
	\footnote{
	With $
		H_0
		=
			73
			~\kilo\metre
			~\reciprocal{\second}
			~\reciprocal{\mega\parsec}
		$,
	the linear Hubble's law gives
		$
		H_0 
		\times
			384,400
			~\kilo\metre
		=
			2.87
			~\centi\metre
			~\reciprocal{\yyear}
		$.
	This is relativistically na\"ive,
	because
		the relativistic model anticipates
			no expansion on short scales
		\cite[p719]{MTW},
	and the gravitational deceleration originally expected
		should have reduced the expansion to virtually vanish
			even at $1~\AU$
		\cite{Cooperstock1998},
	but is correct for the present theory
		in which the recessions are merely apparent.
	The excess fits
		a known mismatch factor of $5$
			with laboratory frictional coefficients
		\cite{Lambeck1977},
	currently attributed to
		previously missing model detail
		\cite{Kagan1997,Ray1999}.
	Since tidal dissipation concerns energy,
	the corrected apparent recession would be
		$
		3.82
		\times \sqrt{5}/(1 + \sqrt{5})
		\approx
			2.64
		$,
	leaving a margin of
		$
		1.18
			~\centi\metre
			~\reciprocal{\yyear}
		$,
		for the real recession,
	closer to historical recession rates
		indicated by paleontological data
	that would not have been affected by
		phase acceleration,
	of about
		$
		1.27
			~\centi\metre
			~\reciprocal{\yyear}
		$
		from
		$
		- 2.5
			~\giga\yyear
		$
		to
		$
		- 650
			~\mega\yyear
		$
		\cite{Williams1990,Williams1997}.
	This mismatch too is currently attributed to
		the lack of small seas in the past.
	Correcting for time dilation due to
		present clock drift
	brings the values even closer in line.
	}, 
again consistent
	the uniform apparent expansion of all measured distances
that would be expected for
	a systematic instrument cause like creep.
More importantly,
a correction for the dilation of past time
	would be required in \emph{all} measurements.
Even NASA's consideration of atomic clock drifts
	in the context of the Pioneer anomaly
has been limited to verification of
	the Allan variances
	\cite[~\S{}V-C]{Anderson2002a},
which are essentially autocorrelations
	computed by local circuits
		that would be similarly subject to creep.
The time dilation correction would increase
	the sun's age
	to over $
	7
		~\giga{\yyear}
	$
and that of the universe, to \emph{infinity},
	consistent with the above inference for nuclear abundances,
and more conservative
	than the standard model.



\section*{Conclusion} 

I have shown that
	a drift of scale in
		frequency or wavelength selection
will cause 
	phase acceleration in proportion to
		the slope of the phase spectrum
			of the received radiation,
and result in
	scaling of subsequently detected spectra
in proportion to
	the source distance of the radiation.
The acceleration must occur because
	the waveform arriving at the subsequent detector
comprises
	instantaneous values of the radiation
		from different successive sinusoids
present in the Fourier decomposition of
	the original waveform.
The spectral scaling must occur since
	the detector can only integrate
		the waveform actually presented to it
	by the preceding selection.
The phase spectral slope to support
	the phase acceleration
requires
	continuity of the received spectrum,
which is in turn guaranteed by
	the finite beginning and end implicit for any wavepacket
		transporting information or energy.
Proportionality to the source distances
is assured by
	the total path delay from the source,
	proportionality of the phase
		represented by the fixed path delay
			in an individual sinusoid
		to its frequency,
and
	the dominance of this path phase contribution
		at large distances
	over any initial phase differences
		between the sinusoids.
All of the mathematical premises are basic and
	independent of relativity or quantum theory,
and the mathematical consistency of the prediction
	has been verified by
		digital simulation of linespreads
			with the path delay
	using
		variable sampling to simulate
			receiver clock drift
equivalent to
	a drift of its frequency or wavelength selection.

This prediction is as yet theoretic,
	like Doppler's in 1842
	\cite{Doppler1842}.
The adequacy of real linespreads,
	for example
		of continuous wave lasers and of radio waves,
for the envisaged
	terrestrial and noncosmological applications
		is yet to be established.
A simulation available online%
	\footnote{
	As a Java applet at
	\texttt{http://www.inspiredresearch.com}.
	This was used for the test results
		presented in Section \ref{s:veri}.
	} 
shows that
	optical devices may soon reach
		the continuous tuning speeds
	necessary for
		a visual laboratory test of the theory,
and also that
	radio frequency tests and applications
		should at most require
			modified RF processing
and would be well in reach of
	existing technology.
The implications for
	astronomical spectroscopy and cosmology
		(Section \ref{s:broad})
concern subcycle drifts of scale
	even over lengthy photon integrations,
and should therefore hold regardless of
	immediate success or failure
		in terrestrial validation.


\section*{Acknowledgements} 

I sincerely thank Bruce Elmegreen
	for years of patient listening and guidance
		in astrophysics,
John Kosinski
	for persistent encouragement,
and
the SPIE session chair, Katherine Creath,
	for crucial help in making this readable.


\appendix
\begin{raggedright}

\end{raggedright}
\end{document}